\documentclass[10pt,
superscriptaddress,
 amsmath,amssymb,
 aps,
prl,
twocolumn,
]{revtex4-2}

\usepackage{braket}
\usepackage{graphicx}
\usepackage[version=4]{mhchem}

\usepackage{cancel,siunitx}
\usepackage{upgreek}

\usepackage[dvipsnames]{xcolor}

\usepackage[caption=false]{subfig} 
\usepackage{ragged2e} 
\DeclareCaptionJustification{justified}{\justifying}
\usepackage[colorlinks=true, allcolors=blue]{hyperref}

\begin{document}

\title{Barium Autoionization for Efficient Ion Trap Loading}
\author{Zachary J. Wall}
\email{zwall@physics.ucla.edu}
\affiliation{Department of Physics and Astronomy, University of California Los Angeles, Los Angeles, CA, USA}

\author{Justin D. Piel}
\affiliation{Department of Physics and Astronomy, Purdue University, West Lafayette, Indiana 47907 USA}

\author{Samuel R. Vizvary}
\affiliation{Department of Physics and Astronomy, University of California Los Angeles, Los Angeles, CA, USA}

\author{Michael Bareian}
\affiliation{Department of Physics and Astronomy, University of California Los Angeles, Los Angeles, CA, USA}

\author{Steven Diaz}
\affiliation{Department of Physics and Astronomy, University of California Los Angeles, Los Angeles, CA, USA}

\author{Elijah Mossman}
\affiliation{Department of Physics and Astronomy, University of California Los Angeles, Los Angeles, CA, USA}

\author{Anthony Ransford}
\affiliation{Quantinuum, LLC, 303 South Technology Court, Broomfield, Colorado 80021, USA}

\author{Chris H. Greene}
\affiliation{Department of Physics and Astronomy, Purdue University, West Lafayette, Indiana 47907 USA}
\author{Eric R. Hudson}
\affiliation{Department of Physics and Astronomy, University of California Los Angeles, Los Angeles, CA, USA}
\affiliation{Challenge Institute for Quantum Computation, University of California Los Angeles, Los Angeles, CA, USA}
\affiliation{Center for Quantum Science and Engineering, University of California Los Angeles, Los Angeles, CA, USA}

\author{Wesley C. Campbell}
\affiliation{Department of Physics and Astronomy, University of California Los Angeles, Los Angeles, CA, USA}
\affiliation{Challenge Institute for Quantum Computation, University of California Los Angeles, Los Angeles, CA, USA}
\affiliation{Center for Quantum Science and Engineering, University of California Los Angeles, Los Angeles, CA, USA}

\date{\today} 
\begin{abstract}
We report a theoretical and experimental investigation of autoionizing resonances from the $5d6p\,{}^3\mathrm{D}_1^o$ manifold in neutral barium for efficient loading of ion traps.
Our calculations predict large resonant cross sections for many narrow autoionizing resonances, but we find experimentally that for most of these, Doppler broadening during trap loading depresses the effective cross sections that can be achieved in practice.  
We identify and demonstrate a strong, broad transition at $531\,\mathrm{nm}$, and show that it furnishes an order-of-magnitude increase in trap loading efficiency compared to other demonstrated resonances.
\end{abstract}

\maketitle
Barium has emerged as a leading trapped-ion qubit due to its visible-wavelength transitions and long lived ($\tau \approx 30 \,\mathrm{s}$) metastable ${}^2\mathrm{D}_{5/2}$ state \cite{Moses2023,Mehdi2025,Ransford2025}. 
These features make it a favorable system for implementing the \textit{omg} (optical-metastable-ground) blueprint that utilizes three qubit encodings in a single atom \cite{Allcock2021,Yang2022}  The long-wavelength cooling transition also reduces the rate of surface charging when using integrated photonic systems \cite{Mordini2025}.  Further, the recently-introduced synthetic ${}^{133}\mathrm{Ba}^+$ isotope incorporates all of these features as well as fast, high-fidelity qubit state preparation \cite{Christensen2020,Hucul2017,Boguslawski2023,Vizvary2024}. However ${}^{133}\mathrm{Ba}^+$ is radioactive, and therefore requires working with small quantities that must be loaded efficiently \cite{Greenberg2023}.  Even for naturally-abundant isotopes, current loading rates can require minutes to populate a moderate-sized processor \cite{Moses2023}, and will prove infeasible for operating much larger systems.  More-efficient loading methods will likely be needed for scaling and may enable wider adoption of trace species that furnish desirable advantages over more-abundant nuclides.

Inspired by the recent work of Greenberg \textit{et al.} \cite{Greenberg2023}, we address this challenge by identifying and characterizing autoionizing resonances in neutral barium with the goal of enhancing ion trap loading efficiency. 
This work provides theoretical calculations and measurements of the cross section of multiple autoionizing resonances from the $5d6p\,{}^3\mathrm{D}_1^o$ state. Specifically, we focus on regions of the ionization spectrum between $450$ and $454 \,\mathrm{nm}$, as well as a broad resonance at $530.8 \,\mathrm{nm}$. The $530.8 \,\mathrm{nm}$ resonance was chosen due to its large predicted cross section and broad spectral width, combined with the availability of high-power lasers at this wavelength.

Previous studies have measured the cross sections and ion trap loading rates from the $6s6p\,{}^1\mathrm{P}_1^o$ \cite{Leschhorn2012,Greenberg2023,White2022,Armstrong1993}, $6s6p\,{}^3\mathrm{P}_1^o$ \cite{Steele2007}, and $5d6p\,{}^3\mathrm{D}_1^o$ \cite{Wood1993} states. 
Here, we provide direct comparison to the most efficient loading rates reported in previous studies and find that the autoionizing transition at $530.8 \,\mathrm{nm}$ is $\approx \! 10$ times more efficient than the best currently known scheme.
A key factor that contributes to the utility of this transition is that the second photon absorption exhibits a linear scaling of loading efficiency with laser power up to extremely high intensities, with no visible saturation effect. 
This is compounded with the practical advantage of high available laser power at this wavelength. 
The transition wavelength is also comfortably in the visible range, which is expected to suppress trap-charging effects that occur when using ultraviolet photoionization light \cite{Harlander_2010,Lucas2004, Qiao2021}. Finally the same photoionization laser can also be used for high fidelity stimulated-Raman-transition gates in $\mathrm{Ba}^+$ \cite{Boguslawski2023}.

\begin{figure}[ht]
    \centering
    \includegraphics[width=8.5cm,height=7.559cm]{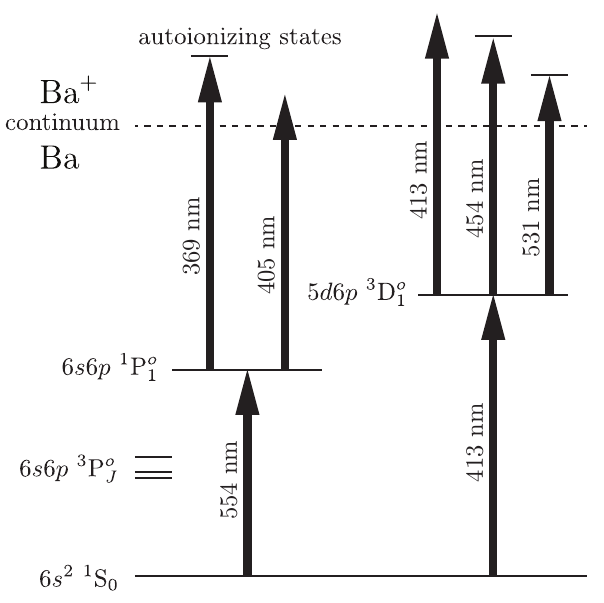}
    \caption{Level diagram of measured photoionization schemes in this work.  The $413\,\mathrm{nm}$ + $531\,\mathrm{nm}$ scheme on the far right is found to be the most efficient for loading barium ions into the trap.}
    \label{fig:Ionization_Schemes}
\end{figure}
\begin{figure*}[t]
\includegraphics[width=.98\textwidth]{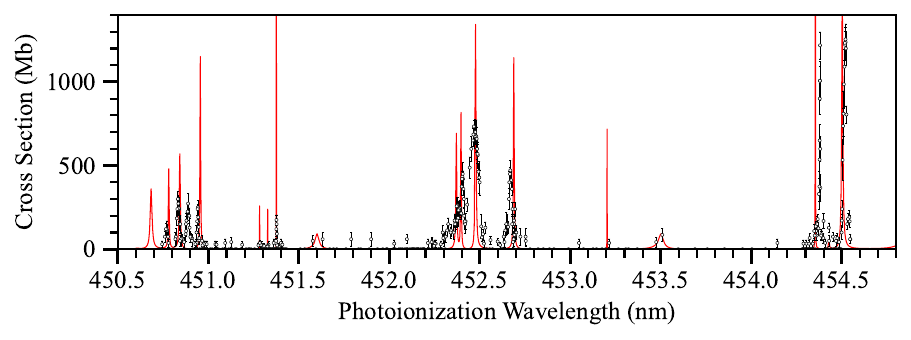}
    \caption{\label{fig:454 Large Range}Spectrum of the autoionizing resonances originating from the $5d6p\,{}^{3}\mathrm{D}_1^o$ state. Data points show the measured cross section, and red lines indicate the theoretical calculations. The calculated peak positions are in reasonable agreement with the measured frequencies, although the peak heights of the narrower resonances exhibit a mismatch.}
    \vspace{-5pt}
\end{figure*}

\begin{figure}[ht]
    \centering
    \includegraphics[width=.49\textwidth]{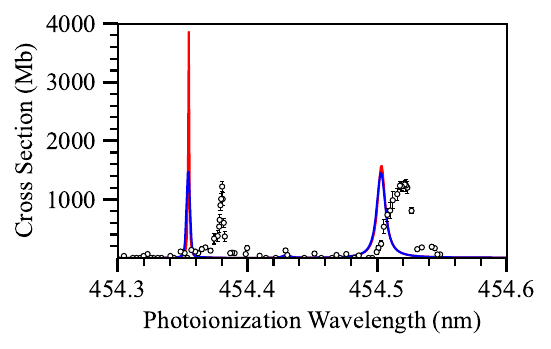}
    \caption{Autoionization resonances to the $5d\,{}_{3/2}15d\,{}_{3/2} \,\,J=2$ state (right) and the $5d\,{}_{3/2}15d\,{}_{5/2} \,\,J=1$ state (left). The disagreement in the height of the autoionizing peaks relative to the theoretical calculations (red) is explained by Doppler broadening, as evidenced by the much closer agreement of the cross section produced by convolving the theoretical curve with a Gaussian profile(blue).}
    \label{fig:455_zoom}
\end{figure}

The theoretical methods used to explore interesting scenarios for creating barium ions from neutrals through two-step photoionization have been well-developed over the past few decades.  The present study adopts the techniques of variational R-matrix theory integrated with multichannel quantum defect theory, along the lines presented in \cite{Aymar1996, Aymar1983}.  The starting point of such calculations for any atom or negative ion with two valence electrons is to first create an accurate model potential energy function $V_{lj}(r)$ that can be used to describe the interaction of a single electron with the closed-shell ionic core, i.e. for one electron of orbital momentum $l$ and total individual angular momentum $j$ in the field of Ba$^{++}$.  This potential is radially local, meaning that for each $l,j$, it is an ordinary function of the distance $r$ between that electron and the nucleus of the closed-shell ion.  This potential is optimized to give the energy levels of Ba$^+$ as accurately as possible for $l<5$.  The physics incorporated in $V_{lj}(r)$ is primarily the screening effects of the closed-shell electrons of the Ba$^{++}$ ion, its polarizability interaction with one electron, and the spin-orbit interaction of that electron in the field of the doubly-charged ion. The dielectronic polarization term has not been incorporated into the present calculations, but from past experience it is not expected to be important in this energy range.

Relative cross sections were experimentally measured by laser ablating a BaCl$_2$ target to produce neutral barium in the trap volume, and the number of loaded ions were counted. We set the experimental cross section equal to the calculated cross section at 413 nm, and scaled all other measured cross sections relative to this value. All isotopes of barium that were loaded into the trap were recorded. A $532 \,\mathrm{nm}$ Nd:YAG, focused to a beam waist of $300 \,\mathrm{\text\textmu m}$ with a pulse energy of $0.5 \,\mathrm{mJ}$ was used for laser ablation. The target was positioned at a distance of $4.3 \,\mathrm{cm}$ from the trap center. Approximately $500 \,\mathrm{\text\textmu W}$ of $413 \,\mathrm{nm}$ light was used for the first step of photoionization ($6s^2\,{}^1\mathrm{S}_0 \rightarrow 5d6p\,{}^3\mathrm{D}_1^o$).  This beam was vertically polarized and overlapped with the $493 \,\mathrm{nm}$ and $650 \,\mathrm{nm}$ ion-Doppler-cooling lasers. These were directed through the trap center perpendicular to the direction of the ablation cloud to minimize the Doppler shift. The spectroscopy laser was propagated antiparallel to the $413 \,\mathrm{nm}$ beam with horizontal polarization.

The ion trap used in this experiment is made up of four, segmented, gold-coated ceramic blades. The ion-to-electrode distance is $200 \,\mathrm{\text\textmu m}$ with a trap drive frequency of  $\Omega_\mathrm{RF} = 2\pi \times 32.0$ MHz. The RF voltage is set to an amplitude of $200 \,\mathrm{V}$ resulting in a secular frequency of $\omega_\mathrm{radial} = 2\pi \times 2.0\,\mathrm{MHz}$ and a radial trap depth of $4.3 \,\mathrm{eV}$. Axial confinement is achieved by applying $20\,\mathrm{V}$ of static bias to the four outermost electrodes, resulting in an axial secular frequency of $\omega_\mathrm{axial} = 2\pi \times 0.55$ MHz. The trap is housed in a sealed vacuum chamber with a measured pressure of approximately $2 \times 10^{-10} \,\mathrm{Torr}$.

The first photoionization scheme newly studied in this work begins with a resonant transition at 413 nm from the $6s^2\,{}^1\mathrm{S}_0$ state to the $5d6p\,{}^3\mathrm{D}_1^o$ state. A second spectroscopy laser is then scanned from $450.5 \,\mathrm{nm}$ to $454.8\,\mathrm{nm}$, driving to various autoionizing states. The full range of the spectroscopy is shown in Fig.~\ref{fig:454 Large Range}.

A key objective of this spectroscopy was to validate that the theoretically calculated spectrum aligns with experimental measurements in both its frequency distribution and total cross section. A significant discrepancy was identified upon observing that, although the peak frequencies of the spectroscopy results show close agreement, the cross sections of the narrowest peaks are notably diminished when measured. A plausible explanation for this observation is Doppler broadening of the high-temperature ablation cloud. The temperature of the incoming ablation plume can be approximated by measuring the broadening of the first step of the photoionization and comparing this to the known linewidth of the $6s^2\,{}^1\mathrm{S}_0 \rightarrow 5d6p\,{}^3\mathrm{D}_1^o$  transition. A frequency scan of the $413 \,\mathrm{nm}$ transition was fit to a Voigt profile to determine the Gaussian linewidth resulting from Doppler broadening. A Gaussian with a full width half maximum corresponding to a Doppler width of 350 K is then convolved with the theoretical autoionization spectrum.

Fig.~\ref{fig:455_zoom} shows the two strongest resonances before and after the convolution is performed. The theoretical spectrum with the convolution shows a much closer agreement to the experimental cross section, the effect is especially notable for the narrow peak on the left.  We expect loading methods that utilize colder atomic sources, such as magneto-optical trap loading, to yield stronger and less broadened resonances \cite{Johansen2022}.

\begin{figure}[ht]
\includegraphics[width=.49\textwidth]{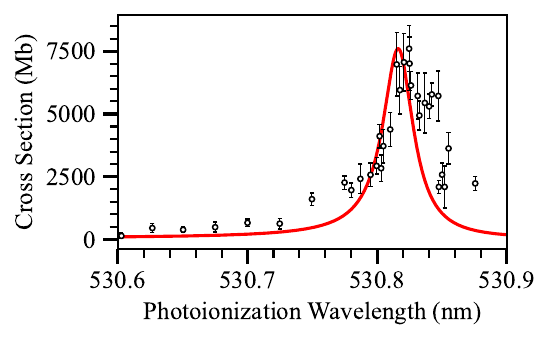}
\caption{\label{fig:531 Spectrum} Spectrum of the autoionizing resonances from the $5d6p\,{}^3\mathrm{D}_1^o$ state. The measured spectrum (black points) agrees well with the theoretical prediction (red lines), without the application of additional broadening.}
\vspace{-5pt}
\end{figure}

After surveying the full theoretical spectrum, taking into account the Doppler broadening effects, a strong resonance driving to the $5d\,{}_{3/2}8d\,{}_{5/2} \,\,J=2$ state at $530.8 \,\mathrm{nm}$  was identified as the ideal peak for photoionization. This transition exhibits the highest  cross section of the measured spectrum, and occurs at a wavelength where current laser technology can deliver extremely high power. Precision spectroscopy of this transition was performed with a $1.1 \,\mathrm{W}$
fiber laser system, shown in  Fig.~\ref{fig:531 Spectrum}.  The cross section is experimentally confirmed to be $(7400 \pm  870)\,\mathrm{Mb}$, which is consistent with the theoretical calculation within the experimental standard deviation.

A third, previously studied \cite{Greenberg2023}, photoionization scheme was measured to compare its efficiency with that of the newly characterized schemes. This scheme drives the resonant 554 nm transition to the $6s6p\,{}^3\mathrm{P}_1^o$  state followed by non-resonant excitation to the ionization continuum by a $405\, \mathrm{nm}$ laser. This scheme was chosen to provide a direct comparison to the work in \cite{Greenberg2023} which characterized both this scheme and the resonant autoionizing scheme utilizing a $389\, \mathrm{nm}$ transition, coupling to the $5d\,{}_{5/2}8d\,{}_{5/2}\,\,J=1$. The cross section at $405 \,\mathrm{nm}$ was measured to be $(72 \pm 3)\,\mathrm{Mb}$, in good agreement with the theoretical value of $76 \,\mathrm{Mb}$.  The cross section of the autoionizing $389 \,\mathrm{nm}$ transition was measured in \cite{Greenberg2023} to be approximately 7 times more efficient than this non-resonant $405 \,\mathrm{nm}$ photoionization scheme. This differs from the theoretically calculated cross section which would suggest a factor of 17. 

\begin{figure}[ht]
\includegraphics[width=.49\textwidth]{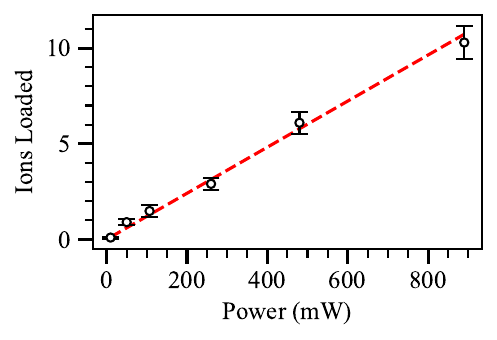}
\caption{\label{fig:Power_vs_ions}Average number of ions loaded per ablation shot as a function of 530.8 nm laser power. The red dashed line shows a linear fit to the data ($\chi^2/\mathrm{dof} = 1.01$), indicating that the loading remains linear over the explored power range with no evidence of saturation.}
\vspace{-5pt}
\end{figure}

It is important to consider the practical utility as well as the cross section when comparing the known photoionization pathways. An advantage of the newly developed $531 \,\mathrm{nm}$ photoionization scheme is the abundance of laser power at this wavelength which significantly enhances ion trap loading efficiency. This high laser power can be leveraged to linearly increase the loading rate up to an extremely high saturation limit. The saturation limit of this transition was calculated to be $.86 \,\mathrm{MW}/\mathrm{cm}^2$, owing to the fact that the lifetime of the autoionizing state is on the order of picoseconds. Fig.~\ref{fig:Power_vs_ions} shows the linear relationship between the input laser intensity and ion loading rate per ablation shot.

\begin{figure}[ht]
\includegraphics[width=.49\textwidth]{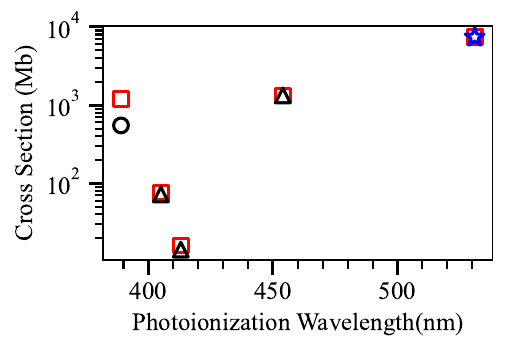}
\caption{\label{fig:Comparison_Chart} Comparison of  relevant photoionization cross sections. The blue star shows the cross section of the 531 nm transition. Black triangles are other measured loading rates in this work, the black circle is the measured loading rate in 
\cite{Greenberg2023}, driving from the $6s6p\,{}^3\mathrm{P}_1^o$ state. Red squares are theoretical calculations from this work.}

\vspace{-5pt}
\end{figure}

In summary, two novel autoionization pathways with high cross sections have been measured. Scheme 1, utilizing the autoionizing $454 \,\mathrm{nm}$ transition, demonstrates a method that is roughly equal in photoionization efficiency and available laser power to the previously known most efficient $389 \,\mathrm{nm}$ autoionization scheme. Scheme 2, drives the resonant $531 \,\mathrm{nm}$ transition and shows approximately an order of magnitude increase in cross section over the $389 \,\mathrm{nm}$ scheme. A comparison of the cross section of all relevant photoionization pathways considered in this work is shown in Fig.~\ref{fig:Comparison_Chart}.
The higher cross section of Scheme 2 is coupled with the practical advantage of multiple orders of magnitude of increased available laser power, which directly translates to higher ionization rates. Additionally, the $531 \,\mathrm{nm}$ transition, being in the visible spectrum, mitigates anomalous charging effects that are commonly observed with ultraviolet photoionization lasers.  These capabilities are especially valuable for loading radioactive isotopes like ${}^{133}\mathrm{Ba}^+$, where maximizing ion yield is essential for overcoming material scarcity and radioactivity constraints.  By leveraging this transition, future ion trap experiments can achieve significantly faster and more reliable ion loading, reducing downtime and improving scalability for quantum technologies.

\bigskip

\begin{acknowledgments}
This work was supported in part by the Army Research Office grant W911NF-20-1-0037 and National Science Foundation grants PHY-2207985, PHY-2207546, and OMA-2016245.  The theory work at Purdue University by CHG and JDP was supported in part by an AFOSR-MURI Grant (No. FA9550-20-1-
0323). We thank Quantinuum for supplying the $554 \,\mathrm{nm}$ laser used in this work.
\end{acknowledgments}

\bibliographystyle{apsrev4-2}
\bibliography{ref}

\begin{thebibliography}{24}%
\makeatletter
\providecommand \@ifxundefined [1]{%
 \@ifx{#1\undefined}
}%
\providecommand \@ifnum [1]{%
 \ifnum #1\expandafter \@firstoftwo
 \else \expandafter \@secondoftwo
 \fi
}%
\providecommand \@ifx [1]{%
 \ifx #1\expandafter \@firstoftwo
 \else \expandafter \@secondoftwo
 \fi
}%
\providecommand \natexlab [1]{#1}%
\providecommand \enquote  [1]{``#1''}%
\providecommand \bibnamefont  [1]{#1}%
\providecommand \bibfnamefont [1]{#1}%
\providecommand \citenamefont [1]{#1}%
\providecommand \href@noop [0]{\@secondoftwo}%
\providecommand \href [0]{\begingroup \@sanitize@url \@href}%
\providecommand \@href[1]{\@@startlink{#1}\@@href}%
\providecommand \@@href[1]{\endgroup#1\@@endlink}%
\providecommand \@sanitize@url [0]{\catcode `\\12\catcode `\$12\catcode
  `\&12\catcode `\#12\catcode `\^12\catcode `\_12\catcode `\%12\relax}%
\providecommand \@@startlink[1]{}%
\providecommand \@@endlink[0]{}%
\providecommand \url  [0]{\begingroup\@sanitize@url \@url }%
\providecommand \@url [1]{\endgroup\@href {#1}{\urlprefix }}%
\providecommand \urlprefix  [0]{URL }%
\providecommand \Eprint [0]{\href }%
\providecommand \doibase [0]{https://doi.org/}%
\providecommand \selectlanguage [0]{\@gobble}%
\providecommand \bibinfo  [0]{\@secondoftwo}%
\providecommand \bibfield  [0]{\@secondoftwo}%
\providecommand \translation [1]{[#1]}%
\providecommand \BibitemOpen [0]{}%
\providecommand \bibitemStop [0]{}%
\providecommand \bibitemNoStop [0]{.\EOS\space}%
\providecommand \EOS [0]{\spacefactor3000\relax}%
\providecommand \BibitemShut  [1]{\csname bibitem#1\endcsname}%
\let\auto@bib@innerbib\@empty
\bibitem [{\citenamefont {Moses}\ \emph {et~al.}(2023)\citenamefont {Moses}
  \emph {et~al.}}]{Moses2023}%
  \BibitemOpen
  \bibfield  {author} {\bibinfo {author} {\bibfnamefont {S.~A.}\ \bibnamefont
  {Moses}} \emph {et~al.},\ }\href {https://doi.org/10.1103/PhysRevX.13.041052}
  {\bibfield  {journal} {\bibinfo  {journal} {Phys. Rev. X}\ }\textbf {\bibinfo
  {volume} {13}},\ \bibinfo {pages} {041052} (\bibinfo {year}
  {2023})}\BibitemShut {NoStop}%
\bibitem [{\citenamefont {Mehdi}\ \emph {et~al.}(2025)\citenamefont {Mehdi},
  \citenamefont {Vaidya}, \citenamefont {Savill-Brown}, \citenamefont
  {Grosser}, \citenamefont {Ratcliffe}, \citenamefont {Liu}, \citenamefont
  {Haine}, \citenamefont {Hope},\ and\ \citenamefont {Viteri}}]{Mehdi2025}%
  \BibitemOpen
  \bibfield  {author} {\bibinfo {author} {\bibfnamefont {Z.}~\bibnamefont
  {Mehdi}}, \bibinfo {author} {\bibfnamefont {V.~D.}\ \bibnamefont {Vaidya}},
  \bibinfo {author} {\bibfnamefont {I.}~\bibnamefont {Savill-Brown}}, \bibinfo
  {author} {\bibfnamefont {P.}~\bibnamefont {Grosser}}, \bibinfo {author}
  {\bibfnamefont {A.~K.}\ \bibnamefont {Ratcliffe}}, \bibinfo {author}
  {\bibfnamefont {H.}~\bibnamefont {Liu}}, \bibinfo {author} {\bibfnamefont
  {S.~A.}\ \bibnamefont {Haine}}, \bibinfo {author} {\bibfnamefont {J.~J.}\
  \bibnamefont {Hope}},\ and\ \bibinfo {author} {\bibfnamefont {C.~R.}\
  \bibnamefont {Viteri}},\ }\href {https://arxiv.org/pdf/2412.07185} {\bibfield
   {journal} {\bibinfo  {journal} {Phys. Rev. Lett.}\ } (\bibinfo {year}
  {2025})}\BibitemShut {NoStop}%
\bibitem [{\citenamefont {Ransford}\ \emph {et~al.}(2025)\citenamefont
  {Ransford}, \citenamefont {Allman}, \citenamefont {Arkinstall}, \citenamefont
  {Campora} \emph {et~al.}}]{Ransford2025}%
  \BibitemOpen
  \bibfield  {author} {\bibinfo {author} {\bibfnamefont {A.}~\bibnamefont
  {Ransford}}, \bibinfo {author} {\bibfnamefont {M.~S.}\ \bibnamefont
  {Allman}}, \bibinfo {author} {\bibfnamefont {J.}~\bibnamefont {Arkinstall}},
  \bibinfo {author} {\bibfnamefont {J.~P.}\ \bibnamefont {Campora}}, \emph
  {et~al.}\ }\href {https://doi.org/10.48550/arXiv.2511.05465}
  {10.48550/arXiv.2511.05465} (\bibinfo {year} {2025}),\ \Eprint
  {https://arxiv.org/abs/2511.05465} {arXiv:2511.05465 [quant-ph]} \BibitemShut
  {NoStop}%
\bibitem [{\citenamefont {Allcock}\ \emph {et~al.}(2021)\citenamefont
  {Allcock}, \citenamefont {Campbell}, \citenamefont {Chiaverini},
  \citenamefont {Chuang}, \citenamefont {Hudson}, \citenamefont {Moore},
  \citenamefont {Ransford}, \citenamefont {Roman}, \citenamefont {Sage},\ and\
  \citenamefont {Wineland}}]{Allcock2021}%
  \BibitemOpen
  \bibfield  {author} {\bibinfo {author} {\bibfnamefont {D.~T.~C.}\
  \bibnamefont {Allcock}}, \bibinfo {author} {\bibfnamefont {W.~C.}\
  \bibnamefont {Campbell}}, \bibinfo {author} {\bibfnamefont {J.}~\bibnamefont
  {Chiaverini}}, \bibinfo {author} {\bibfnamefont {I.~L.}\ \bibnamefont
  {Chuang}}, \bibinfo {author} {\bibfnamefont {E.~R.}\ \bibnamefont {Hudson}},
  \bibinfo {author} {\bibfnamefont {I.~D.}\ \bibnamefont {Moore}}, \bibinfo
  {author} {\bibfnamefont {A.}~\bibnamefont {Ransford}}, \bibinfo {author}
  {\bibfnamefont {C.}~\bibnamefont {Roman}}, \bibinfo {author} {\bibfnamefont
  {J.~M.}\ \bibnamefont {Sage}},\ and\ \bibinfo {author} {\bibfnamefont
  {D.~J.}\ \bibnamefont {Wineland}},\ }\href
  {https://doi.org/10.1063/5.0069544} {\bibfield  {journal} {\bibinfo
  {journal} {Appl. Phys. Lett.}\ }\textbf {\bibinfo {volume} {119}},\ \bibinfo
  {pages} {214002} (\bibinfo {year} {2021})}\BibitemShut {NoStop}%
\bibitem [{\citenamefont {Yang}\ \emph {et~al.}(2022)\citenamefont {Yang},
  \citenamefont {Ma}, \citenamefont {Wu}, \citenamefont {Wang}, \citenamefont
  {Cao}, \citenamefont {Guo}, \citenamefont {Huang}, \citenamefont {Feng},
  \citenamefont {Zhou},\ and\ \citenamefont {Duan}}]{Yang2022}%
  \BibitemOpen
  \bibfield  {author} {\bibinfo {author} {\bibfnamefont {H.-X.}\ \bibnamefont
  {Yang}}, \bibinfo {author} {\bibfnamefont {J.-Y.}\ \bibnamefont {Ma}},
  \bibinfo {author} {\bibfnamefont {Y.-K.}\ \bibnamefont {Wu}}, \bibinfo
  {author} {\bibfnamefont {Y.}~\bibnamefont {Wang}}, \bibinfo {author}
  {\bibfnamefont {M.-M.}\ \bibnamefont {Cao}}, \bibinfo {author} {\bibfnamefont
  {W.-X.}\ \bibnamefont {Guo}}, \bibinfo {author} {\bibfnamefont {Y.-Y.}\
  \bibnamefont {Huang}}, \bibinfo {author} {\bibfnamefont {L.}~\bibnamefont
  {Feng}}, \bibinfo {author} {\bibfnamefont {Z.-C.}\ \bibnamefont {Zhou}},\
  and\ \bibinfo {author} {\bibfnamefont {L.-M.}\ \bibnamefont {Duan}},\ }\href
  {https://doi.org/https://doi.org/10.1038/s41567-022-01661-5} {\bibfield
  {journal} {\bibinfo  {journal} {Nat. Phys}\ }\textbf {\bibinfo {volume}
  {18}},\ \bibinfo {pages} {1058} (\bibinfo {year} {2022})}\BibitemShut
  {NoStop}%
\bibitem [{\citenamefont {Mordini}\ \emph {et~al.}(2025)\citenamefont
  {Mordini}, \citenamefont {Ricci~Vasquez}, \citenamefont {Motohashi},
  \citenamefont {M\"uller}, \citenamefont {Malinowski}, \citenamefont {Zhang},
  \citenamefont {Mehta}, \citenamefont {Kienzler},\ and\ \citenamefont
  {Home}}]{Mordini2025}%
  \BibitemOpen
  \bibfield  {author} {\bibinfo {author} {\bibfnamefont {C.}~\bibnamefont
  {Mordini}}, \bibinfo {author} {\bibfnamefont {A.}~\bibnamefont
  {Ricci~Vasquez}}, \bibinfo {author} {\bibfnamefont {Y.}~\bibnamefont
  {Motohashi}}, \bibinfo {author} {\bibfnamefont {M.}~\bibnamefont {M\"uller}},
  \bibinfo {author} {\bibfnamefont {M.}~\bibnamefont {Malinowski}}, \bibinfo
  {author} {\bibfnamefont {C.}~\bibnamefont {Zhang}}, \bibinfo {author}
  {\bibfnamefont {K.~K.}\ \bibnamefont {Mehta}}, \bibinfo {author}
  {\bibfnamefont {D.}~\bibnamefont {Kienzler}},\ and\ \bibinfo {author}
  {\bibfnamefont {J.~P.}\ \bibnamefont {Home}},\ }\href
  {https://doi.org/10.1103/PhysRevX.15.011040} {\bibfield  {journal} {\bibinfo
  {journal} {Phys. Rev. X}\ }\textbf {\bibinfo {volume} {15}},\ \bibinfo
  {pages} {011040} (\bibinfo {year} {2025})}\BibitemShut {NoStop}%
\bibitem [{\citenamefont {Christensen}\ \emph {et~al.}(2020)\citenamefont
  {Christensen}, \citenamefont {Hucul}, \citenamefont {Hudson},\ and\
  \citenamefont {Campbell}}]{Christensen2020}%
  \BibitemOpen
  \bibfield  {author} {\bibinfo {author} {\bibfnamefont {J.~E.}\ \bibnamefont
  {Christensen}}, \bibinfo {author} {\bibfnamefont {D.}~\bibnamefont {Hucul}},
  \bibinfo {author} {\bibfnamefont {E.~R.}\ \bibnamefont {Hudson}},\ and\
  \bibinfo {author} {\bibfnamefont {W.~C.}\ \bibnamefont {Campbell}},\
  }\bibfield  {journal} {\bibinfo  {journal} {npj Quantum Information}\
  }\textbf {\bibinfo {volume} {6}},\ \href
  {https://doi.org/10.1038/s41534-020-0265-5} {10.1038/s41534-020-0265-5}
  (\bibinfo {year} {2020})\BibitemShut {NoStop}%
\bibitem [{\citenamefont {Hucul}\ \emph {et~al.}(2017)\citenamefont {Hucul},
  \citenamefont {Christensen}, \citenamefont {Hudson},\ and\ \citenamefont
  {Campbell}}]{Hucul2017}%
  \BibitemOpen
  \bibfield  {author} {\bibinfo {author} {\bibfnamefont {D.}~\bibnamefont
  {Hucul}}, \bibinfo {author} {\bibfnamefont {J.~E.}\ \bibnamefont
  {Christensen}}, \bibinfo {author} {\bibfnamefont {E.~R.}\ \bibnamefont
  {Hudson}},\ and\ \bibinfo {author} {\bibfnamefont {W.~C.}\ \bibnamefont
  {Campbell}},\ }\href {https://doi.org/10.1103/PhysRevLett.119.100501}
  {\bibfield  {journal} {\bibinfo  {journal} {Phys. Rev. Lett.}\ }\textbf
  {\bibinfo {volume} {119}},\ \bibinfo {pages} {100501} (\bibinfo {year}
  {2017})}\BibitemShut {NoStop}%
\bibitem [{\citenamefont {Boguslawski}\ \emph {et~al.}(2023)\citenamefont
  {Boguslawski}, \citenamefont {Wall}, \citenamefont {Vizvary}, \citenamefont
  {Moore}, \citenamefont {Bareian}, \citenamefont {Allcock}, \citenamefont
  {Wineland}, \citenamefont {Hudson},\ and\ \citenamefont
  {Campbell}}]{Boguslawski2023}%
  \BibitemOpen
  \bibfield  {author} {\bibinfo {author} {\bibfnamefont {M.~J.}\ \bibnamefont
  {Boguslawski}}, \bibinfo {author} {\bibfnamefont {Z.~J.}\ \bibnamefont
  {Wall}}, \bibinfo {author} {\bibfnamefont {S.~R.}\ \bibnamefont {Vizvary}},
  \bibinfo {author} {\bibfnamefont {I.~D.}\ \bibnamefont {Moore}}, \bibinfo
  {author} {\bibfnamefont {M.}~\bibnamefont {Bareian}}, \bibinfo {author}
  {\bibfnamefont {D.~T.~C.}\ \bibnamefont {Allcock}}, \bibinfo {author}
  {\bibfnamefont {D.~J.}\ \bibnamefont {Wineland}}, \bibinfo {author}
  {\bibfnamefont {E.~R.}\ \bibnamefont {Hudson}},\ and\ \bibinfo {author}
  {\bibfnamefont {W.~C.}\ \bibnamefont {Campbell}},\ }\href
  {https://doi.org/10.1103/PhysRevLett.131.063001} {\bibfield  {journal}
  {\bibinfo  {journal} {Phys. Rev. Lett.}\ }\textbf {\bibinfo {volume} {131}},\
  \bibinfo {pages} {063001} (\bibinfo {year} {2023})}\BibitemShut {NoStop}%
\bibitem [{\citenamefont {Vizvary}\ \emph {et~al.}(2024)\citenamefont
  {Vizvary}, \citenamefont {Wall}, \citenamefont {Boguslawski}, \citenamefont
  {Bareian}, \citenamefont {Derevianko}, \citenamefont {Campbell},\ and\
  \citenamefont {Hudson}}]{Vizvary2024}%
  \BibitemOpen
  \bibfield  {author} {\bibinfo {author} {\bibfnamefont {S.~R.}\ \bibnamefont
  {Vizvary}}, \bibinfo {author} {\bibfnamefont {Z.~J.}\ \bibnamefont {Wall}},
  \bibinfo {author} {\bibfnamefont {M.~J.}\ \bibnamefont {Boguslawski}},
  \bibinfo {author} {\bibfnamefont {M.}~\bibnamefont {Bareian}}, \bibinfo
  {author} {\bibfnamefont {A.}~\bibnamefont {Derevianko}}, \bibinfo {author}
  {\bibfnamefont {W.~C.}\ \bibnamefont {Campbell}},\ and\ \bibinfo {author}
  {\bibfnamefont {E.~R.}\ \bibnamefont {Hudson}},\ }\href
  {https://doi.org/10.1103/PhysRevLett.132.263201} {\bibfield  {journal}
  {\bibinfo  {journal} {Phys. Rev. Lett.}\ }\textbf {\bibinfo {volume} {132}},\
  \bibinfo {pages} {263201} (\bibinfo {year} {2024})}\BibitemShut {NoStop}%
\bibitem [{\citenamefont {N.~Greenberg}\ and\ \citenamefont
  {Senko}(2024)}]{Greenberg2023}%
  \BibitemOpen
  \bibfield  {author} {\bibinfo {author} {\bibfnamefont {P.~J.~L.}\
  \bibnamefont {N.~Greenberg}, \bibfnamefont {B.~M.~White}}\ and\ \bibinfo
  {author} {\bibfnamefont {C.}~\bibnamefont {Senko}},\ }\href
  {https://doi.org/10.1088/2058-9565/ad3f41} {\bibfield  {journal} {\bibinfo
  {journal} {Quantum Sci. Technol}\ }\textbf {\bibinfo {volume} {9}},\ \bibinfo
  {pages} {035023} (\bibinfo {year} {2024})}\BibitemShut {NoStop}%
\bibitem [{\citenamefont {Leschhorn}\ \emph {et~al.}(2012)\citenamefont
  {Leschhorn}, \citenamefont {Hasegawa},\ and\ \citenamefont
  {Schaetz}}]{Leschhorn2012}%
  \BibitemOpen
  \bibfield  {author} {\bibinfo {author} {\bibfnamefont {G.}~\bibnamefont
  {Leschhorn}}, \bibinfo {author} {\bibfnamefont {T.}~\bibnamefont
  {Hasegawa}},\ and\ \bibinfo {author} {\bibfnamefont {T.}~\bibnamefont
  {Schaetz}},\ }\href {https://doi.org/10.1007/s00340-012-5101-y} {\bibfield
  {journal} {\bibinfo  {journal} {Appl Phys B}\ }\textbf {\bibinfo {volume}
  {108}},\ \bibinfo {pages} {159} (\bibinfo {year} {2012})}\BibitemShut
  {NoStop}%
\bibitem [{\citenamefont {White}\ \emph {et~al.}(2022)\citenamefont {White},
  \citenamefont {Low}, \citenamefont {de~Sereville}, \citenamefont {Day},
  \citenamefont {Greenberg}, \citenamefont {Rademacher},\ and\ \citenamefont
  {Senko}}]{White2022}%
  \BibitemOpen
  \bibfield  {author} {\bibinfo {author} {\bibfnamefont {B.~M.}\ \bibnamefont
  {White}}, \bibinfo {author} {\bibfnamefont {P.~J.}\ \bibnamefont {Low}},
  \bibinfo {author} {\bibfnamefont {Y.}~\bibnamefont {de~Sereville}}, \bibinfo
  {author} {\bibfnamefont {M.~L.}\ \bibnamefont {Day}}, \bibinfo {author}
  {\bibfnamefont {N.}~\bibnamefont {Greenberg}}, \bibinfo {author}
  {\bibfnamefont {R.}~\bibnamefont {Rademacher}},\ and\ \bibinfo {author}
  {\bibfnamefont {C.}~\bibnamefont {Senko}},\ }\href
  {https://doi.org/10.1103/PhysRevA.105.033102} {\bibfield  {journal} {\bibinfo
   {journal} {Phys. Rev. A}\ }\textbf {\bibinfo {volume} {105}},\ \bibinfo
  {pages} {033102} (\bibinfo {year} {2022})}\BibitemShut {NoStop}%
\bibitem [{\citenamefont {Armstrong}\ \emph {et~al.}(1993)\citenamefont
  {Armstrong}, \citenamefont {Wood},\ and\ \citenamefont
  {Greene}}]{Armstrong1993}%
  \BibitemOpen
  \bibfield  {author} {\bibinfo {author} {\bibfnamefont {D.~J.}\ \bibnamefont
  {Armstrong}}, \bibinfo {author} {\bibfnamefont {R.~P.}\ \bibnamefont
  {Wood}},\ and\ \bibinfo {author} {\bibfnamefont {C.~H.}\ \bibnamefont
  {Greene}},\ }\href {https://doi.org/10.1103/PhysRevA.47.1981} {\bibfield
  {journal} {\bibinfo  {journal} {Phys. Rev. A}\ }\textbf {\bibinfo {volume}
  {47}},\ \bibinfo {pages} {1981} (\bibinfo {year} {1993})}\BibitemShut
  {NoStop}%
\bibitem [{\citenamefont {Steele}\ \emph {et~al.}(2007)\citenamefont {Steele},
  \citenamefont {Churchill}, \citenamefont {Griffin},\ and\ \citenamefont
  {Chapman}}]{Steele2007}%
  \BibitemOpen
  \bibfield  {author} {\bibinfo {author} {\bibfnamefont {A.~V.}\ \bibnamefont
  {Steele}}, \bibinfo {author} {\bibfnamefont {L.~R.}\ \bibnamefont
  {Churchill}}, \bibinfo {author} {\bibfnamefont {P.~F.}\ \bibnamefont
  {Griffin}},\ and\ \bibinfo {author} {\bibfnamefont {M.~S.}\ \bibnamefont
  {Chapman}},\ }\href {https://doi.org/10.1103/PhysRevA.75.053404} {\bibfield
  {journal} {\bibinfo  {journal} {Phys. Rev. A}\ }\textbf {\bibinfo {volume}
  {75}},\ \bibinfo {pages} {053404} (\bibinfo {year} {2007})}\BibitemShut
  {NoStop}%
\bibitem [{\citenamefont {Wood}\ \emph {et~al.}(1993)\citenamefont {Wood},
  \citenamefont {Greene},\ and\ \citenamefont {Armstrong}}]{Wood1993}%
  \BibitemOpen
  \bibfield  {author} {\bibinfo {author} {\bibfnamefont {R.~P.}\ \bibnamefont
  {Wood}}, \bibinfo {author} {\bibfnamefont {C.~H.}\ \bibnamefont {Greene}},\
  and\ \bibinfo {author} {\bibfnamefont {D.}~\bibnamefont {Armstrong}},\ }\href
  {https://doi.org/10.1103/PhysRevA.47.229} {\bibfield  {journal} {\bibinfo
  {journal} {Phys. Rev. A}\ }\textbf {\bibinfo {volume} {47}},\ \bibinfo
  {pages} {229} (\bibinfo {year} {1993})}\BibitemShut {NoStop}%
\bibitem [{\citenamefont {Harlander}\ \emph {et~al.}(2010)\citenamefont
  {Harlander}, \citenamefont {Brownnutt}, \citenamefont {Hänsel},\ and\
  \citenamefont {Blatt}}]{Harlander_2010}%
  \BibitemOpen
  \bibfield  {author} {\bibinfo {author} {\bibfnamefont {M.}~\bibnamefont
  {Harlander}}, \bibinfo {author} {\bibfnamefont {M.}~\bibnamefont
  {Brownnutt}}, \bibinfo {author} {\bibfnamefont {W.}~\bibnamefont {Hänsel}},\
  and\ \bibinfo {author} {\bibfnamefont {R.}~\bibnamefont {Blatt}},\ }\href
  {https://doi.org/10.1088/1367-2630/12/9/093035} {\bibfield  {journal}
  {\bibinfo  {journal} {New Journal of Physics}\ }\textbf {\bibinfo {volume}
  {12}},\ \bibinfo {pages} {093035} (\bibinfo {year} {2010})}\BibitemShut
  {NoStop}%
\bibitem [{\citenamefont {Lucas}\ \emph {et~al.}(2004)\citenamefont {Lucas},
  \citenamefont {Ramos}, \citenamefont {Home}, \citenamefont {McDonnell},
  \citenamefont {Nakayama}, \citenamefont {Stacey}, \citenamefont {Webster},
  \citenamefont {Stacey},\ and\ \citenamefont {Steane}}]{Lucas2004}%
  \BibitemOpen
  \bibfield  {author} {\bibinfo {author} {\bibfnamefont {D.~M.}\ \bibnamefont
  {Lucas}}, \bibinfo {author} {\bibfnamefont {A.}~\bibnamefont {Ramos}},
  \bibinfo {author} {\bibfnamefont {J.~P.}\ \bibnamefont {Home}}, \bibinfo
  {author} {\bibfnamefont {M.~J.}\ \bibnamefont {McDonnell}}, \bibinfo {author}
  {\bibfnamefont {S.}~\bibnamefont {Nakayama}}, \bibinfo {author}
  {\bibfnamefont {J.-P.}\ \bibnamefont {Stacey}}, \bibinfo {author}
  {\bibfnamefont {S.~C.}\ \bibnamefont {Webster}}, \bibinfo {author}
  {\bibfnamefont {D.~N.}\ \bibnamefont {Stacey}},\ and\ \bibinfo {author}
  {\bibfnamefont {A.~M.}\ \bibnamefont {Steane}},\ }\href
  {https://doi.org/10.1103/PhysRevA.69.012711} {\bibfield  {journal} {\bibinfo
  {journal} {Phys. Rev. A}\ }\textbf {\bibinfo {volume} {69}},\ \bibinfo
  {pages} {012711} (\bibinfo {year} {2004})}\BibitemShut {NoStop}%
\bibitem [{\citenamefont {Qiao}\ \emph {et~al.}(2021)\citenamefont {Qiao},
  \citenamefont {Wang}, \citenamefont {Cai}, \citenamefont {Du}, \citenamefont
  {Wang}, \citenamefont {Luan}, \citenamefont {Chen}, \citenamefont {Noh},\
  and\ \citenamefont {Kim}}]{Qiao2021}%
  \BibitemOpen
  \bibfield  {author} {\bibinfo {author} {\bibfnamefont {M.}~\bibnamefont
  {Qiao}}, \bibinfo {author} {\bibfnamefont {Y.}~\bibnamefont {Wang}}, \bibinfo
  {author} {\bibfnamefont {Z.}~\bibnamefont {Cai}}, \bibinfo {author}
  {\bibfnamefont {B.}~\bibnamefont {Du}}, \bibinfo {author} {\bibfnamefont
  {P.}~\bibnamefont {Wang}}, \bibinfo {author} {\bibfnamefont {C.}~\bibnamefont
  {Luan}}, \bibinfo {author} {\bibfnamefont {W.}~\bibnamefont {Chen}}, \bibinfo
  {author} {\bibfnamefont {H.-R.}\ \bibnamefont {Noh}},\ and\ \bibinfo {author}
  {\bibfnamefont {K.}~\bibnamefont {Kim}},\ }\href
  {https://doi.org/10.1103/PhysRevLett.126.023604} {\bibfield  {journal}
  {\bibinfo  {journal} {Phys. Rev. Lett.}\ }\textbf {\bibinfo {volume} {126}},\
  \bibinfo {pages} {023604} (\bibinfo {year} {2021})}\BibitemShut {NoStop}%
\bibitem [{\citenamefont {Aymar}\ \emph {et~al.}(1996)\citenamefont {Aymar},
  \citenamefont {Greene},\ and\ \citenamefont {Luc-Koenig}}]{Aymar1996}%
  \BibitemOpen
  \bibfield  {author} {\bibinfo {author} {\bibfnamefont {M.}~\bibnamefont
  {Aymar}}, \bibinfo {author} {\bibfnamefont {C.~H.}\ \bibnamefont {Greene}},\
  and\ \bibinfo {author} {\bibfnamefont {E.}~\bibnamefont {Luc-Koenig}},\
  }\href {https://doi.org/10.1103/RevModPhys.68.1015} {\bibfield  {journal}
  {\bibinfo  {journal} {Reviews of Modern Physics}\ }\textbf {\bibinfo {volume}
  {68}},\ \bibinfo {pages} {1015} (\bibinfo {year} {1996})}\BibitemShut
  {NoStop}%
\bibitem [{\citenamefont {M.~Aymar}\ and\ \citenamefont
  {Himdy}(1983)}]{Aymar1983}%
  \BibitemOpen
  \bibfield  {author} {\bibinfo {author} {\bibfnamefont {P.~C.}\ \bibnamefont
  {M.~Aymar}}\ and\ \bibinfo {author} {\bibfnamefont {A.~E.}\ \bibnamefont
  {Himdy}},\ }\bibfield  {journal} {\bibinfo  {journal} {Phys. Scr.}\ }\textbf
  {\bibinfo {volume} {193}},\ \href
  {https://doi.org/10.1088/0031-8949/27/3/007} {10.1088/0031-8949/27/3/007}
  (\bibinfo {year} {1983})\BibitemShut {NoStop}%
\bibitem [{\citenamefont {Johansen}\ \emph {et~al.}(2022)\citenamefont
  {Johansen}, \citenamefont {Estey}, \citenamefont {Rowe},\ and\ \citenamefont
  {Ransford}}]{Johansen2022}%
  \BibitemOpen
  \bibfield  {author} {\bibinfo {author} {\bibfnamefont {J.}~\bibnamefont
  {Johansen}}, \bibinfo {author} {\bibfnamefont {B.}~\bibnamefont {Estey}},
  \bibinfo {author} {\bibfnamefont {M.}~\bibnamefont {Rowe}},\ and\ \bibinfo
  {author} {\bibfnamefont {A.}~\bibnamefont {Ransford}},\ }in\ \href
  {https://doi.org/10.1109/QCE53715.2022.00050} {\emph {\bibinfo {booktitle}
  {2022 IEEE International Conference on Quantum Computing and Engineering
  (QCE)}}}\ (\bibinfo {year} {2022})\ pp.\ \bibinfo {pages}
  {299--303}\BibitemShut {NoStop}%
\bibitem [{\citenamefont {Choi}\ \emph {et~al.}(2024)\citenamefont {Choi},
  \citenamefont {Lee}, \citenamefont {Yum}, \citenamefont {An},\ and\
  \citenamefont {Kim}}]{Choi2024}%
  \BibitemOpen
  \bibfield  {author} {\bibinfo {author} {\bibfnamefont {J.}~\bibnamefont
  {Choi}}, \bibinfo {author} {\bibfnamefont {E.}~\bibnamefont {Lee}}, \bibinfo
  {author} {\bibfnamefont {D.}~\bibnamefont {Yum}}, \bibinfo {author}
  {\bibfnamefont {K.}~\bibnamefont {An}},\ and\ \bibinfo {author}
  {\bibfnamefont {J.}~\bibnamefont {Kim}},\ }\href
  {https://doi.org/10.1103/PhysRevA.110.032812} {\bibfield  {journal} {\bibinfo
   {journal} {Phys. Rev. A}\ }\textbf {\bibinfo {volume} {110}},\ \bibinfo
  {pages} {032812} (\bibinfo {year} {2024})}\BibitemShut {NoStop}%
\bibitem [{\citenamefont {Fano}\ and\ \citenamefont {Rau}(1986)}]{FanoRau1986}%
  \BibitemOpen
  \bibfield  {author} {\bibinfo {author} {\bibfnamefont {U.}~\bibnamefont
  {Fano}}\ and\ \bibinfo {author} {\bibfnamefont {A.~R.~P.}\ \bibnamefont
  {Rau}},\ }\href@noop {} {\emph {\bibinfo {title} {Atomic Collisions and
  Spectra}}}\ (\bibinfo  {publisher} {Academic Press},\ \bibinfo {address}
  {Orlando},\ \bibinfo {year} {1986})\BibitemShut {NoStop}%
\end{thebibliography}%
\clearpage

\onecolumngrid      

\appendix
\section{Supplementary Information}

\section{\MakeLowercase{6s}\ce{^{1}_{}S_0} to \MakeLowercase{5d6p}\ce{^{3}_{}D_1} Linewidth Measurement}

\begin{figure}[ht]
\includegraphics[width=.49\textwidth]{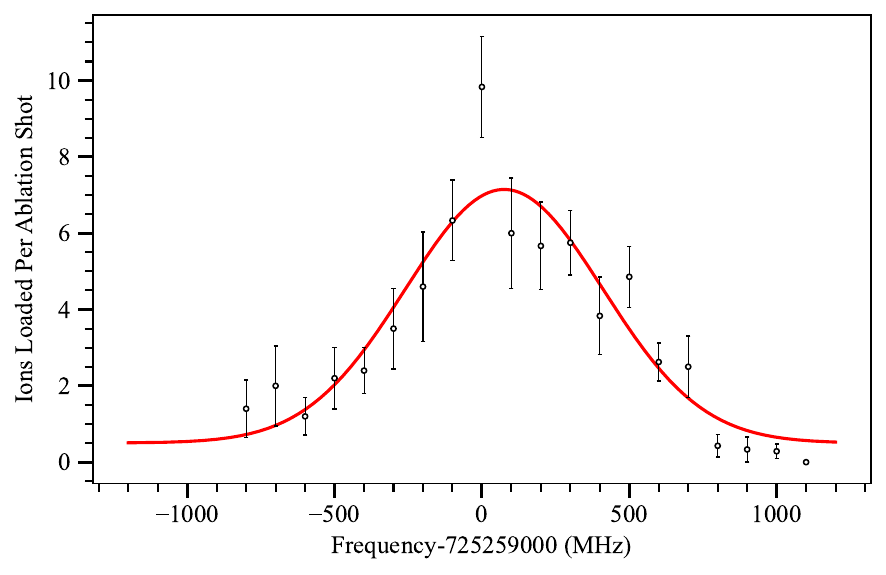}
    \caption{\label{fig:413_Voight}A frequency scan of the 413 nm transition was fit with a Voigt profile, from which a Gaussian component of $\sigma = 335 $ MHz was extracted. This doppler width corresponds to a temperature of 318 K at the center of the trap. Time of flight data was also taken at this ablation fluence and used to measure an ion velocity at the trap center of 205 m/s, corresponding to a temperature at ablation of 350 K. The center frequency of 725529000 (MHz) agrees with the previously reported spectroscopy in \cite{Choi2024}}
    \vspace{-5pt}
\end{figure}

\section{Calculation of Saturation Intensity From Cross Section}
We can extract the dipole moment for each resonance in theoretical spectrum by calculating the oscillator strength. First we convert from cross section $\sigma$ in units of Megabarns to $\frac{df}{de}$ in units of $eV^{-1}$.
We use equation 2.18 \cite{FanoRau1986}, to do this.
$$\sigma = 2\pi^2e^2\frac{\hbar}{mc}\frac{df}{d\epsilon}$$
With this we can find $\frac{df}{de}$ at the relevant peak. The converted spectrum is fit  to a Lorentzian $f(x)$ to get an oscillator strength and linewidth.
$$f(x) = f_s \frac{1}{\pi}\frac{\gamma}{(x-x_0)^2+\gamma^2}$$
This yields an oscilator strength of $f_s = 0.01051$ and a linewidth of $\gamma =0.1329$ eV.
Now we can use equation 2.23 from \cite{FanoRau1986} to extract the dipole element from the oscillator strength
$$f_s = (2m\omega_s/\hbar)|\bra{s}x\ket{0}|^2$$
or solving for the dipole moment
$$|\bra{s}x\ket{0}| = \sqrt{\hbar f_s/(2m\omega_s)}$$
Plugging in for $f_s$ and noting $\omega_s=2\pi\times563.7\times 10^{12}$ we get 
$$|\bra{s}ex\ket{0}| = 2.05\times 10^{-30}$$
Next, with a fitted linewidth of 32.1 GHz, for a 100 $\mu m$ beam waist we can calculate saturation power of $67.43$ Watts.

$$P_{Sat} \text{ @ 531 nm}  = 67.43 W$$

\end{document}